\documentclass[aps, superscriptaddress,amsmath,amssymb,floatfix,bm,braket,notitlepage,twocolumn,prx]{revtex4-1}
\usepackage{times}
\usepackage{graphicx}
\usepackage{color}
\usepackage{bm}
\usepackage{braket}
\usepackage{mathtools}
\usepackage{mathcomp}
\usepackage[caption=false,justification=raggedright,position=top,singlelinecheck=off]{subfig}

\usepackage{soul}

\begin{document}

\title{Orbital selectivity and emergent superconducting state from 
quasi-degenerate $s-$ and $d-$wave 
pairing channels in iron-based superconductors}

\author{Emilian M. Nica}
\email[Corresponding author: ]{en5@rice.edu}
\affiliation{Department of Physics and Astronomy, Rice University, Houston,
Texas 77005}

\author{Rong Yu}
\affiliation{Department of Physics, Renmin University of China, Beijing 100872, China}

\author{Qimiao Si}
\affiliation{Department of Physics and Astronomy, Rice University, Houston,
Texas 77005}

\date{\today}

\begin{abstract}
A major puzzle about the nature of the iron-based superconductivity appears in the case of the alkaline iron selenides. Compared to the iron pnictides, these systems have only electron Fermi pockets ({\it i.e.}, 
no hole Fermi pockets) but comparable superconducting transition temperatures. 
The challenge lies in reconciling the two basic experimental features 
of their superconducting state: a node-less gap and the 
existence of a resonance in the spin excitation spectrum. Here we propose a mechanism based on 
reconstructing two quasi-degenerate pairing states, 
one in an $s$-wave $A_{1g}$ channel that is fully gapped, and the other in a $d$-wave $B_{1g}$ 
channel whose pairing function changes sign across the electron Fermi pockets at the Brillouin-zone boundary. 
The resulting intermediate pairing state,
which we call an orbital-selective $s \times \tau_3$ state,
incorporates both of the above two properties. 
When the leading spin-singlet pairing is in the $d_{xz}, d_{yz}$ orbital subspace, 
this pairing state retains the 
$s$-wave form factor but has a $B_{1g}$ symmetry due to an internal $\tau_3$ 
structure in the orbital space. 
Within a five-orbital $t-J_{1}-J_{2}$ model with orbital-selective exchange couplings, 
we show that the proposed pairing state is energetically competitive over a finite range of control parameters. 
We calculate the dynamical spin susceptibility in the orbital-selective $s \times \tau_3$  superconducting state and show that a spin resonance 
arises and has the characteristics of observed by inelastic neutron experiments in the alkaline iron selenides. 
More generally, the formation of the orbital-selective $s \times \tau_3$ state represents a novel means of relieving the quasi-degeneracy between $s-$ and $d-$wave pairing states, 
which is a hitherto unsuspected alternative to the conventional route of linearly superposing the two into a time-reversal symmetry breaking $s+id$ state.
\end{abstract}

\maketitle

\section{Introduction }
\label{Sec:Space_group_no_SOC }

Understanding the nature of the iron-based superconductivity remains a central challenge 
in condensed matter physics. The superconductivity grows out of a bad-metal normal state, 
with room-temperature resistivity that reaches the Mott-Ioffe-Regel limit \cite{Stewart:Rev_Mod_Phys_2001, Qazilbash:Nat_Phys_2009}. 
This observation has been interpreted in terms of electron correlations that are sufficiently strong to place the system in proximity to
 an electronic localization transition \cite{Si_Abrahams:PRL_2008,Yin_Haule_Kotliar:Nat_Phys_2011}. 
In addition, the iron-based superconductivity typically occurs near an antiferromagnetic ground state \cite{Dai:arxiv_2015}. 
The combination of these features have motivated a strong-coupling approach \cite{Yu_Nat_Comm:2013,Seo2008,Goswami_Nikolic_Si:EPL_2010}, 
in which short-range exchange interactions among quasi-local moments 
drive the formation of Cooper pairs. Minimal exchange couplings involve $J_1$, 
the interaction among the nearest-neighbor sites on the Fe-square lattice, and $J_2$, 
its next-nearest-neighbor counterpart. Inelastic neutron scattering experiments have demonstrated the importance of such interactions, both for the iron pnictides and iron chalcogenides \cite{Dai:arxiv_2015}. 
The $J_2$ interaction promotes an $s-$wave $A_{1g}$ state, for which there is considerable evidence in the iron pnictides case \cite{Stewart:Rev_Mod_Phys_2001}. In this scenario, the pairing wave function 
changes sign between the hole Fermi surfaces near the center of the Brillouin zone and 
the electron Fermi surfaces at the boundary of the Brillouin zone \cite{Wang_Lee:Science_2011}, 
a switch which is believed to be responsible for a resonance spin excitation at the wave vector $(\pi,0)$ \cite{Dai:arxiv_2015}. 
For the properties of the superconducting state, the distinction
between the strong and weak coupling 
approaches arise subtly. For instance, 
in the strong coupling approach, the pairing order parameter is naturally defined in the orbital basis,
such that the local Coulomb repulsion is minimized \cite{Anderson:Science_2007}, ultimately leading to the orbital selectivity of the gap function \cite{Yu_Zhu_Si:2014, Yin:Nat_Phys_2014}. The latter, in turn, can cause the appearance of double resonances in the spin excitation spectrum \cite{Yu_Zhu_Si:2014} as has been experimentally observed \cite{Zhang_et_al:PRL_2014, Zhang_et_al:PRB_2015}.
 
In spite of some theoretical successes, our understanding of the iron-based superconductivity remains quite limited. One way to make
progress is to take advantage of the large materials basis and gain new insights from systems with
different microscopic electronic behavior. In this context, a 
prominent puzzle has come from
the "122" alkaline iron selenide compounds such as
 $ \textnormal{K}_{\textnormal{y}}\textnormal{Fe}_{\textnormal{2-x}}\textnormal{Se}_{\textnormal{2}} $. 
These systems show electron Fermi pockets only, lacking the hole pockets that occur in the iron pnictides at the center of their 1-Fe Brillouin Zone (BZ) \cite{Mou_et_al:PRL_2011, Zhang_et_al:PRL_2014, Wang_et_al:EPL_2011}.
Yet, the superconducting transition temperatures ($T_c$) for the two classes of materials are comparable. 
What is striking is the apparently conflicting nature of the pairing function.
ARPES experiments indicate a fully gapped quasiparticle dispersion \cite{Mou_et_al:PRL_2011, Zhang_et_al:PRL_2014, Wang_et_al:EPL_2011}, including the electron Fermi 
pockets at the center of the BZ \cite{Xu_et_al:PRB_2012,Wang_et_al:EPL_2012}.
This is compatible with the usual $s$-wave $A_{1g}$ pairing state, but not with a $d$-wave $B_{1g}$ state.
On the other hand,
neutron scattering experiments \cite{Park_etal:2011,Friemel_etal:2012}
observe a sharp resonance peak around the
wavevector $(\pi, \pi/2)$.
These observations are
 consistent with a sign change between the two Fermi pockets at the edge of the 1-Fe BZ, such as would occur in a $d$-wave $B_{1g}$ state, but not in the usual $s$-wave $A_{1g}$ case.

In this paper, we identify an intermediate pairing state that is reconstructed from the conventional $s$-wave $A_{1g}$ and $d$-wave $B_{1g}$ states in the parameter regime where the two are quasi-degenerate \cite{Yu_Nat_Comm:2013}. Orbital selectivity plays an essential role 
in  the formation of this state.
When the spin-singlet pairing is restricted to the $d_{xz}, d_{yz}$ orbital subspace,
the candidate pairing state belongs to the $B_{1g}$ representation of the associated point group
but has a form factor belonging to the $A_{1g}$ representation.
Specifically, the pairing function can be written as $\Delta_{0} \times g_{x^2y^2}(\pmb{k}) \times \tau_{3}$, where $g_{x^2y^2}(\pmb{k})=\text{cos}(k_{x})\text{cos}(k_{y})$ is a an s-wave form factor and $\tau_{3}$ is a Pauli matrix in the $2 \times 2, d_{xz}, d_{yz}$ orbital subspace.
Therefore, we'll refer to this state as the orbital-selective $s \times \tau_3$ state.
When the spin-singlet pairings in the remaining orbital sectors are considered, additional $B_{1g}$ components are mixed in but the gap remains node-less. The gap function also changes sign across the electron Fermi pockets at the boundary of the BZ  consequently producing a spin resonance that is consistent with  observation by inelastic neutron experiments on the alkaline iron selenides.

It is instructive to note that the orbital selectivity in the superconducting state can be naturally considered in the case of the multi-orbital iron-based superconductors. Indeed, extensive studies of the salient features of the orbital selectivity in the normal state of these systems have already been carried out, both in theory and experiment \cite{Yu_Si:2011, Yu_Zhu_Si:PRL_2013, deMedici_Giovanetti_Capone:PRL_2014}. 
A particularly striking phenomenon is the orbitally selective Mott transition, where the orbital-dependency is developed to such degree as to allow the vanishing of the spectral weights for a subset of $d$ orbitals. In our discussion below, we will invoke orbital selectivity at two levels. At the level of effective Hamiltonian, we will allow the short-range exchange interactions to be orbital dependent. In addition, at the level of the resulting phases, we will consider pairing functions that are orbitally selective \cite{Yu_Zhu_Si:2014, Yin:Nat_Phys_2014}.

The remainder of the paper is organized
 as follows: Section \ref{Sec:Two_orbital} discusses the $s_{x^2y^2} \times \tau_{3}, B_{1g}$ pairing at length for a simplified two-orbital $d_{xz}, d_{yz}$ system and shows the node-less resulting state and the sign-changing features in the band basis. Section \ref{Sec:Five_orb_Hamilt} introduces the five-orbital $t-J_{1}-J_{2}$ model with orbital-differentiated exchange couplings and comments on the numerical solution. In Section \ref{Sec:Numerical_results}, we show the results obtained from the calculation and highlight the survival of the essential features of the simplified two-orbital case in the more realistic five-orbital systems. Concluding remarks and a summary of the results are presented in Section \ref{Sec:Conclusion}. Some supplementary material is relegated to the Appendix.

\section{The orbital selective $s \times \tau_{3}$ pairing in a two-orbital $d_{xz},d_{yz}$ system}
\label{Sec:Two_orbital}

\begin{figure}[!htb]
\includegraphics[width=0.9\columnwidth]{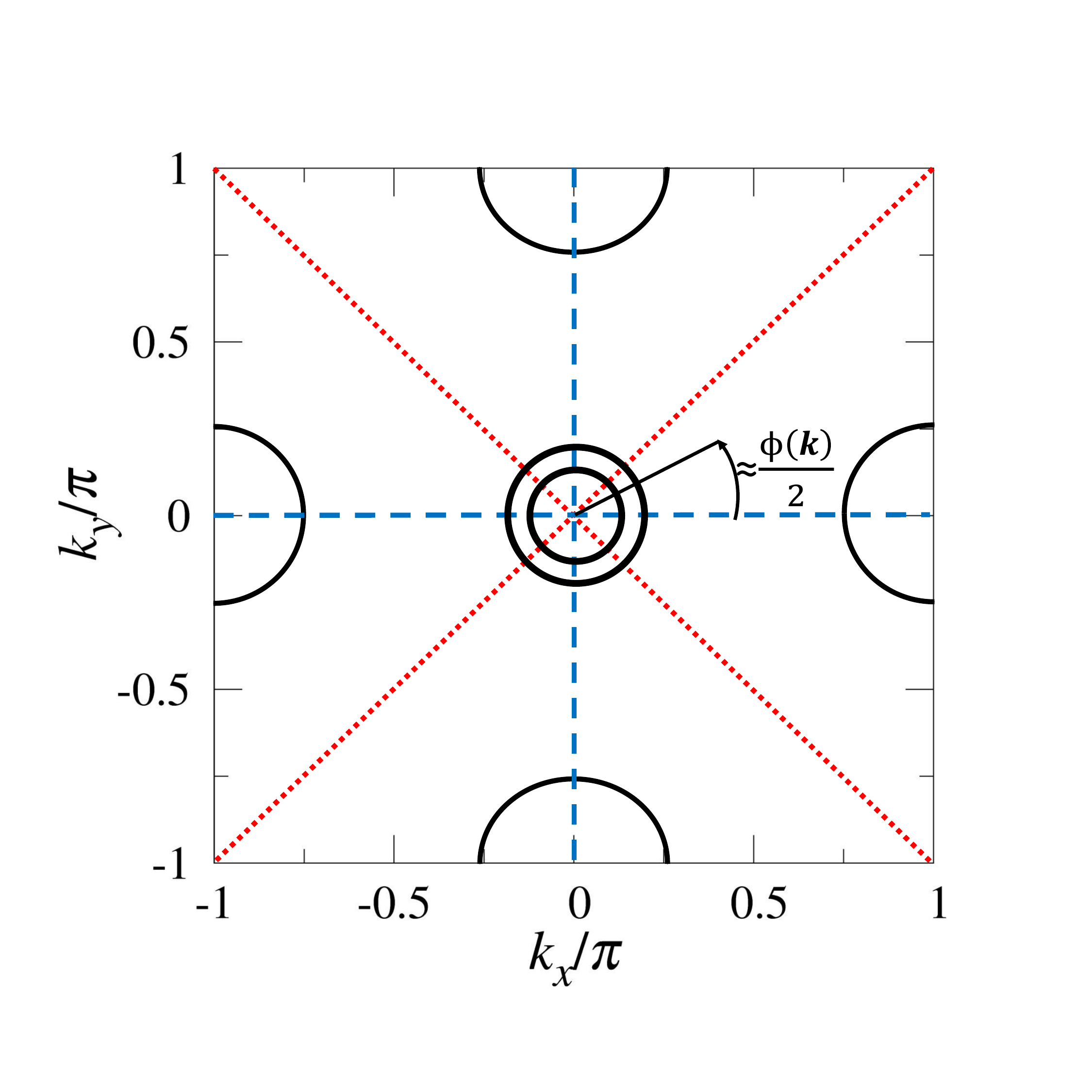}
\caption{\emph{Schematic} illustration of the two-orbital $s_{x^2y^2}, B_{1g}$ pairing in a 1-Fe Brillouin Zone (BZ). The solid lines indicate typical Fermi pockets for the Fe-based superconductors. Both intra- and inter-band components of the pairing shown in  Eq. \ref{Eq:Pairing_band} vanish at $k_{x}=\pm \frac{\pi}{2}$ , $k_{y}=\pm \frac{\pi}{2}$ (not shown) due to the common $g_{x^2y^2}$ form factor. The dotted, red lines indicate the zeroes specific to the band-diagonal pairing (~$\xi_{-}$) while the dashed, blue lines mark the zeroes specific to the off-diagonal band pairing (~$\xi_{xy}$). The typical Fermi surface for a variety of Fe-based superconductors does not extend to the $k_{x}=\pm \frac{\pi}{2}$ , $k_{y}=\pm \frac{\pi}{2}$ common lines of zero pairing. For these types of Fermi surfaces, the intra- and inter- band components do not vanish at the same subset of $\pmb{k}$, ensuring there is always a non-zero pairing given by either of the two components at the 
pockets centered on the origin and edges of the BZ. As explained in the text (Eq. \ref{Eq:Balian_Wert_dispersion}), the two components generate an \emph{effective} gap given by $ \left| d(\pmb{k}) \right| \sim g_{x^2y^2} (\pmb{k})$ with corrections from terms proportional to $\text{sin}^2\phi(\pmb{k})$ (Eqs. \ref{Eq:Balian_Wert_dispersion}-\ref{Eq:Ang_dep} ).  For $\text{max}(\xi_{-}) \approx \text{max}(\xi_{xy})$ the angle $\phi(\pmb{k})$ given by Eq.\ref{Eq:Winding_angle} can be roughly identified with twice the winding angle shown in the figure for fixed $|\pmb{k}|$. The correction from these additional terms does not close the gap,  but does introduce anisotropy in the former. All these make the gap non-zero along a typical Fermi surface for the Fe-based superconductors.  In addition, there is a sign change between the intra-band pairing along the two pockets at the edge of the BZ, a condition necessary to the formation of a resonance in the spin spectrum at the wavevector $\pmb{q}=(\pi, \pi/2)$ observed in experiment \cite{Dai:arxiv_2015}.}
\label{Fig:Empty_BZ}
\end{figure}

We start by introducing the intermediate $s_{x^2y^2} \times \tau_{3}, {B_{1g}}$ spin-singlet pairing
 state in a simplified two-orbital $d_{xz}$, $d_{yz}$ case. Our aim here is to demonstrate that the 
 quasiparticle excitations in this state are fully gapped and, at the same time, that the pairing wave function
 changes sign across the electron Fermi pockets near the M points at the boundary of the {1-Fe} BZ.
To be definite, we discuss the pairing state using a single-particle dispersion for such a two-orbital system
 \cite{Raghu_et_al:2008}.
 Readers not interested in the details of our exposition of the two-orbital case can consult Fig.\ref{Fig:Empty_BZ}, which gives a summary of the most important results of this section.
 
 \noindent 

The Hamiltonian in the orbital basis is given by

\noindent \begin{align}
\label{Eq;Hamilt_orbital}
\hat{H}=\sum_{\pmb{k}} \psi^{\dagger}(\pmb{k}) & \bigg\{ \left[ \xi_{+}(\pmb{k}) \tau_{0} + \xi_{-}(\pmb{k}) \tau_{3} + \xi_{xy}(\pmb{k}) \tau_{1} \right] \otimes \gamma_{3} \notag \\
& + \Delta_{0} g_{x^2y^2}(\pmb{k}) \tau_{3} \otimes \gamma_{1} \bigg\} \psi(\pmb{k}),
\end{align}

\noindent where $\psi^{\dagger}(\pmb{k})=(c^{\dagger}_{\pmb{k} i \sigma}, c_{-\pmb{k} j \sigma'} (i\sigma_{2})_{\sigma' \sigma}  )$ is a spinor in Nambu space, $i,j$ are orbital indices, and $\tau$, $\sigma$ and $\gamma$  are Pauli matrices in the $2 \times 2$ orbital, spin, and Nambu spaces respectively. The exact forms of of the $\xi_{+}, \xi_{-}$, and $\xi_{xy}$ functions and of the resulting bands are given in Appendix \ref{Sec:Appendix}. 
The form of the spinor is chosen to reproduce the antisymmetry of the spin-singlet pairing matrix under exchange. We choose a real amplitude $\Delta_{0}$ for convenience. 

The pairing  changes sign under a $C_{4}$ rotation and belongs to a $B_{1g}$ representation of the associated point group. The minus sign comes entirely from the matrix structure since the transformation leaves the s-wave form factor $g_{x^2y^2}$ invariant.

In order to understand the appearance of a full gap in this $B_{1g}$ pairing we can exploit the analogy
 between the two-orbital system and $^{3}\text{He}$ with spin-triplet pairing. 
 As in Ref.~\onlinecite{Ong_Coleman_Schmalian:2014}, we can define an isospin quantum number 
 for the $2 \times 2$ orbital space and recast the Hamiltonian (Eq.\ref{Eq;Hamilt_orbital}) in Balian-Werthamer form:

\noindent \begin{align}
\label{Eq:Balian_Wert_form}
\hat{H}=\sum_{\pmb{k}} & \psi^{\dagger}_{\pmb{k}} \bigg[ \left( \xi_{+}(\pmb{k}) \tau_{0} +\overrightarrow{B}_{\pmb{k}} \cdot \overrightarrow{\tau} \right) \otimes \gamma_{3} \notag \\
& + \left( \overrightarrow{d}_{\pmb{k}} \cdot \overrightarrow{\tau} \right) \otimes  \gamma_{1} \bigg] \psi_{\pmb{k}}.
\end{align}

\noindent where

\noindent \begin{align}
\overrightarrow{B}(\pmb{k})= & \left( \xi_{xy}(\pmb{k}), 0, \xi_{-}(\pmb{k}) \right)
\nonumber
 \\
\overrightarrow{d}(\pmb{k})= &  \left( 0, 0, \Delta_{0} g_{x^2y^2}(\pmb{k}) \right)
 \label{Eq:Pairing_form}.
\end{align}

As observed in Ref. \onlinecite{Ong_Coleman_Schmalian:2014}, the $\overrightarrow{B}(\pmb{k})$ factor is analogous to a $\pmb{k}$-dependent spin-orbit coupling for $^{3}\text{He}$. We mention some differences. For $^{3}\text{He}$, the spin-orbit coupling forces the spin and spatial degrees of freedom to lock, ensuring that the pairing transform under an element of the point-group $g$ as \cite{Sigrist_Ueda:1991}

\noindent \begin{equation}
\label{Eq:Transformation}
g \overrightarrow{d}(\pmb{k})=\hat{D}^{+}_{G}(g) \overrightarrow{d}( \hat{D}^{-}_{G}(g) \pmb{k}).
\nonumber
\end{equation}

\noindent The matrices $\hat{D}^{+}_{G}(g)$ and $\hat{D}^{-}_{G}(g)$ belong to irreducible representations of the point group even and odd under inversion respectively. In the Fe-based superconductors the  transformation in Eq. \ref{Eq:Transformation} holds even when there is no isospin-orbit coupling $\overrightarrow{B}(\pmb{k})$ because the spatial and orbital degrees of freedom are always locked. In addition, spin-singlet pairing does not force $\overrightarrow{d}(\pmb{k})$ to be odd under space inversion.

To see the consequences of the non-trivial structure of the pairing in orbital space, we write the square of the Hamiltonian matrix:

\noindent \begin{align}
\label{Eq:Squared_Hamilt}
\hat{H}^2= & \sum_{\pmb{k}} \left[ \xi_{+}(\pmb{k}) \tau_{0} + \left( \overrightarrow{B}_{\pmb{k}} \cdot \overrightarrow{\tau} \right) \right]^2 \otimes \gamma_{0} + \left| \overrightarrow{d}(\pmb{k}) \right|^2 \tau_{0} \otimes \gamma_{0} \notag \\
 & + 2i \left( \overrightarrow{B}(\pmb{k}) \times \overrightarrow{d}(\pmb{k}) \right) \cdot \overrightarrow{\tau} \otimes i \gamma_{2}.
\end{align}

\noindent The first two terms are the squares of the free particle Hamiltonian and of a pairing contribution with no structure in orbital space.  The last term is a consequence of the non-commuting free particle and pairing parts of the Hamiltonian. When the commutator is zero this term vanishes while the other two terms reduce to a simple BCS-like matrix, where the anisotropy in the resulting dispersion is entirely determined by the symmetry of the form factor. For example, a $d_{x^2-y^2} \times \tau_{0}, B_{1g}$ pairing will generate nodes along the $k_{x}=k_{y}$ lines in the BZ whereas a $s_{x^2y^2} \times \tau_{0}, A_{1g}$ produces nodes at $k_{x/y}=\pm \pi/2$ in the 1-Fe BZ.  As is apparent from Eq. \ref{Eq:Squared_Hamilt}, the non-Abelian term effectively mixes two BCS-like states and can generate a fully gapped dispersion even when the pairing is of $B_{1g}$ type.

Since the Nambu matrices $\gamma_{0}$ and $i \gamma_{2}$ commute, $\hat{H}^{2}$ in Eq. \ref{Eq:Squared_Hamilt} can be trivially brought to block diagonal form. The quasi-particle dispersion is given by

\noindent
\begin{widetext}
\begin{align}
\label{Eq:Balian_Wert_dispersion}
E_{\pm}(\pmb{k}) &= \sqrt{ \xi^{2}_{+}(\pmb{k}) + \left| \overrightarrow{B}(\pmb{k}) \right|^2 +  \left| \overrightarrow{d}(\pmb{k}) \right|^2  \pm \sqrt{ 4\xi^{2}_{+}(\pmb{k})\left| \overrightarrow{B}(\pmb{k}) \right|^2 + 4\left| \overrightarrow{B}(\pmb{k}) \times \overrightarrow{d}(\pmb{k}) \right|^2  }  } \notag \\
& = \sqrt{ \left ( \sqrt{\xi^{2}_{+}(\pmb{k}) + \left| \overrightarrow{d}(\pmb{k}) \right|^2 \text{sin}^2\phi(\pmb{k}) } \pm \left| \overrightarrow{B}(\pmb{k}) \right|  \right)^2 + \left| \overrightarrow{d}(\pmb{k}) \right|^2 \left( 1- \text{sin}^2\phi(\pmb{k}) \right) }
\end{align}

\noindent \begin{equation}
\label{Eq:Winding_angle}
\text{sin}\phi(\pmb{k})=\frac{\xi_{xy}(\pmb{k})}{\left| \overrightarrow{B}(\pmb{k}) \right|}= \frac{\xi_{xy}(\pmb{k})}{\sqrt{ \xi^2_{-}(\pmb{k})+ \xi^2_{xy}(\pmb{k})}}
\end{equation}

\end{widetext}

\noindent

From Eq. \ref{Eq:Balian_Wert_dispersion} and \emph{general} $\overrightarrow{d}(\pmb{k})$ we see that the quasi-particle dispersion can vanish only when both terms in the square root vanish. The second of these goes to zero when either $\text{sin}\phi(\pmb{k})=1$ or, trivially, when $\left| \overrightarrow{d}(\pmb{k}) \right| = 0$.  When $\text{sin}\phi(\pmb{k})=1$, the dispersion reduces to

\noindent \begin{equation}
\label{Eq:Ang_dep}
E_{\pm}(\pmb{k}) =  \sqrt{\xi^{2}_{+}(\pmb{k}) + \left| \overrightarrow{d}(\pmb{k}) \right|^2 } \pm \left| \overrightarrow{B}(\pmb{k}) \right|,
\end{equation}

\noindent which can acquire \emph{accidental} nodes for general $\pmb{k}$. On the Fermi surface however, we have $\xi^{2}_{+}(\pmb{k})=\left| \overrightarrow{B}(\pmb{k}) \right|^2$ and the expression above is zero only when $\left| \overrightarrow{d}(\pmb{k}) \right| = 0$, that is, along the lines given by the symmetry of the form factor. For our particular choice of $\overrightarrow{d}(\pmb{k})$ (Eq. \ref{Eq:Pairing_form}) corresponding to the $s_{x^2y^2} \times \tau_{3}, B_{1g}$ pairing, these are the $k_{x}=\pm \frac{\pi}{2}$ , $k_{y}=\pm \frac{\pi}{2}$ lines. We conclude that while this type of pairing can generate accidental nodes, it guarantees zero energy states only when the Fermi surface intersects the nodes of the form factor. When the Fermi surface is away from these points, the dispersion will be generally gapped with an enhanced anisotropy relative to the commuting case.

The expression for the quasi-particle dispersion above can also be used to tentatively understand how a pairing of the $s_{x^2y^2} \times \tau_3$ type can become energetically competitive w.r.t. one of a pure $s_{x^2y^2} \times \tau_0$, $d_{x^2-y^2} \times \tau_0$ or even the combined $s+id$ type. In all these cases, the $\left| \overrightarrow{B}(\pmb{k}) \times \overrightarrow{d}(\pmb{k}) \right|^2$ vanishes or equivalently there is no inter-band pairing. The putative gain in energy can be accounted for by the additional anisotropy introduced by the non trivial matrix structure in the orbital space. Although this does not ensure that the $s_{x^2y^2} \times \tau_3$ pairing is always dominant since the minimization of the free-energy generally depends on the specifics of the dispersion and the pairing, the argument above sketches how such a combined state can in principle become leading.

We proceed to comment on another important property of the $s_{x^2y^2} \times \tau_{3} $ pairing function. In the \emph{band} basis the pairing matrix is given by

\noindent \begin{widetext}
\begin{equation}
\label{Eq:Pairing_band}
\hat{\Delta}(\pmb{k})=-\Delta_0 g_{x^2y^2}(\pmb{k}) \left( \frac{\xi_{-}(\pmb{k})}{\sqrt{\xi^2_{-}(\pmb{k})+\xi^2_{xy}(\pmb{k})}} \otimes \alpha_3 + \frac{\xi_{xy}(\pmb{k})}{\sqrt{\xi^2_{-}(\pmb{k})+\xi^2_{xy}(\pmb{k})}} \otimes \alpha_{1} \right)
\end{equation}
\end{widetext}

\noindent where the $\alpha_{1,3}$ are Pauli matrices. Note that there are both diagonal and off-diagonal components which are exactly out of phase. The presence of intra and inter-band pairing makes this situation more appropriate to a strong-coupling system, since both must contribute significantly to the condensation energy.

As can be seen from Eq. \ref{Eq:Pairing_band}, the band-index diagonal term changes sign about the diagonals ($k_x=\pm k_y$) of the BZ while the band-index off-diagonal does so around the axes ($k_{x/y}=0$) of the 1-Fe BZ. For the two orbital system considered in this section, the Fermi pockets at the edge of the 1-Fe BZ are determined by one band alone. We can thus ignore the contribution of the off-diagonal terms ($~\alpha_{1}$ in Eq. \ref{Eq:Pairing_band}) to the spin susceptibility. In our $s_{x^2y^2} \times \tau_{3}$ case, the band diagonal pairing does indeed change sign between the two pockets since $\xi_{-}$ has zeroes along the diagonals of the 1-Fe BZ. This ensures that this type of pairing is conducive to the formation of a resonance (see Appendix \ref{Sec:Appendix} for a review of the typical BCS case).

To summarize, in a two-orbital case, the $s_{x^2y^2} \times \tau_{3}, B_{1g}$ superconducting state has a node-less quasi-particle dispersion with a sign change in the pairing between the two pockets at the 1-Fe BZ edge. 
Unfortunately, the simple features seen in the two-orbital case are not trivially generalizable to the five orbital or even the three-orbital systems since the orbital structure there lacks the simplicity inherent in the Pauli matrices. However, we find that
the key features of the simplified system survive in the full five-orbital scenario.

\section{Orbitally Differentiated Exchange and Pairing in a 5-orbital $t-J_{1}-J_{2}$ Model}
\label{Sec:Five_orb_Hamilt}

We now turn to the question of how the $s_{x^2y^2} \times \tau_{3}, B_{1g}$ pairing state can become energetically competitive. We also partially address the issue of it's relation to the more conventional $s-$ and $d-$wave pairing channels. To make our discussion concrete,  we turn to the five-orbital $t-J_{1}-J_{2}$ model. The use of such a model is motivated by the bad metal nature of the normal state for the iron pnictides as probed by optical conductivity  experiments \cite{Qazilbash:Nat_Phys_2009} and by the the proximity to the insulating state in the case  of the iron selenides. These properties suggest the placement of the two classes of materials in the vicinity of a Mott insulating transition \cite{Yu_Nat_Comm:2013}.

We proceed to describe the effective $t-J_{1}-J_{2}$ model we used in our calculations. These were done for an effective 1-Fe unit cell or equivalently in an unfolded BZ \cite{Nica_Yu_Si:Unpublished_2015}. To simplify our analysis, 
 we consider the free-particle part for all $d$ orbitals but restrict the exchange couplings 
 and hence the pairing interactions to $d_{xz}, d_{yz}$, and $d_{xy}$ orbitals only. 
 Specifically, the Hamiltonian in the orbital basis is given by

\noindent
\begin{widetext}
\begin{align}
\label{Eq:t_J_Hamiltonian}
H=  -  & \sum_{i<j} (t_{ij}^{\alpha\beta}c_{\alpha}^{\dagger}c_{\beta} + H.C.) + \sum_{i,\alpha} \left(  \epsilon_{i\alpha} - \mu \right) n_{i}+  \sum_{<ij>,\alpha, \beta}J_{1}^{\alpha\beta} \left( \pmb{S}_{i\alpha} \cdot \pmb{S_{j\beta}} - \frac{1}{4}n_{i\alpha}n_{j\beta} \right) +  \notag \\
& + \sum_{<<ij>>,\alpha, \beta}J_{2}^{\alpha\beta} \left( \pmb{S}_{i\alpha} \cdot \pmb{S_{j\beta}} - \frac{1}{4}n_{i\alpha}n_{j\beta} \right)
\end{align}

\noindent \begin{equation}
\label{Eq:Orbital_aniso}
J^{xz/yz}_{1,2} \neq  J^{xy}_{1,2}
\end{equation}
\end{widetext}

\noindent where $\alpha,\beta \in \{1,2,3,4,5 \}$ are orbital indices representing all five $d_{xz}$, $d_{yz}$, $d_{x^2-y^2}$, $d_{xy}$, and $d_{3z^2-r^2}$ orbitals, $\epsilon_{i}$ are the on-site energies, and $\mu$ is the chemical potential. The local moments can be written as $\pmb{S}_{i \alpha}=\sum_{ss'}\frac{1}{2}c_{i \alpha s}^{\dagger}\pmb{\sigma}_{ss'}c_{i \alpha s'}$ in terms of the conduction electrons. We take only intra-orbital exchange ($\alpha=\beta$) and set $J^{x^2-y^2}_{1(2)}=J^{3z^2-r^2}_{1(2)}=0$. We consider general exchange couplings which reflect the possible orbital selectivity by allowing  $J_{xz,xz}=J_{yz,yz} \neq J_{xy,xy}$ (Eq. \ref{Eq:Orbital_aniso}).

The interactions in Eq. \ref{Eq:t_J_Hamiltonian} can be decomposed into nearest-neighbor (NN) and next-nearest neighbor (NNN) singlet pairing terms. The double occupancy constraint can be incorporated in practice through a band renormalization by the doping factor $\delta= \left|\sum_{i,s} n_{i\alpha s}- 2 \right|$. The pairing Hamiltonian can be solved numerically in a 1-Fe unit cell mean-field calculation by varying the exchange couplings. For more details on the method, we refer the reader to Refs. \onlinecite{Yu_Nat_Comm:2013, Yu_Zhu_Si:2014}. Here we also define an exchange orbital anisotropy factor $A_{O}=\frac{J_{1}^{xy}}{J_{1}^{xz/yz}}=\frac{J_{2}^{ xy}}{J_{2}^{xz/yz}}$ and an orbital-independent NN-NNN exchange anisotropy factor $A_{L}=\frac{J_{1}^{\alpha}}{J_{2}^{\alpha}}$ for all three non-zero intra-orbital exchange couplings for $d_{xz}$, $d_{yz}$, and $d_{xy}$.

We also calculate the dynamical spin susceptibility in the superconducting state given by

\noindent \begin{equation}
\label{Eq:RPA_suscept}
\chi(\pmb{q}, i\omega_n)=\sum_{\alpha \beta} \chi_{\alpha \beta}(\pmb{q}, i\omega_n),
\end{equation}

\noindent where

\noindent \begin{align}
\chi_{\alpha \beta}(\pmb{q}, i\omega_n)=\sum_{\gamma} & \left[ \textbf{I} + J(\pmb{q}) \sum_{\delta \mu} \chi_{0, \delta \mu} (\pmb{q}, i\omega_n) \right]^{-1}_{\alpha \gamma} \times \notag \\
& \times \chi_{0, \gamma \beta} (\pmb{q}, i\omega_n),
\nonumber
\end{align}
\noindent \begin{equation}
\chi_{0, \alpha \beta}(\pmb{q}, i\omega_n)=\int_{1}^{1/T} d\tau e^{i \omega_{n} \tau } \left< T \left[ S^{-}_{ \pmb{q}\alpha}(\tau)S^{+}_{ \pmb{-q}\beta}(0) \right] \right>,
\nonumber
\end{equation}

\noindent and,

\noindent \begin{equation}
J(\pmb{q})=\frac{J_1}{2} \left( \text{cos} q_x + \text{cos} q_y \right) + J_2 \text{cos} q_x \text{cos} q_y.
\nonumber
\end{equation}

To explore the zero-temperature superconducting phases corresponding to different classes of Fe-based materials we consider the associated free-electron dispersion for $ \textnormal{K}_{\textnormal{y}}\textnormal{Fe}_{\textnormal{2-x}}\textnormal{Se}_{\textnormal{2}} $, iron pnictides and single-layer FeSe. We subsequently tune the exchange couplings for various NN-NNN and orbital anisotropy ratios ($A_L$ and $A_O$) and determine the real-space pairing functions.

\noindent \begin{figure}[t!]
\subfloat[] {\includegraphics[width=0.8\columnwidth]{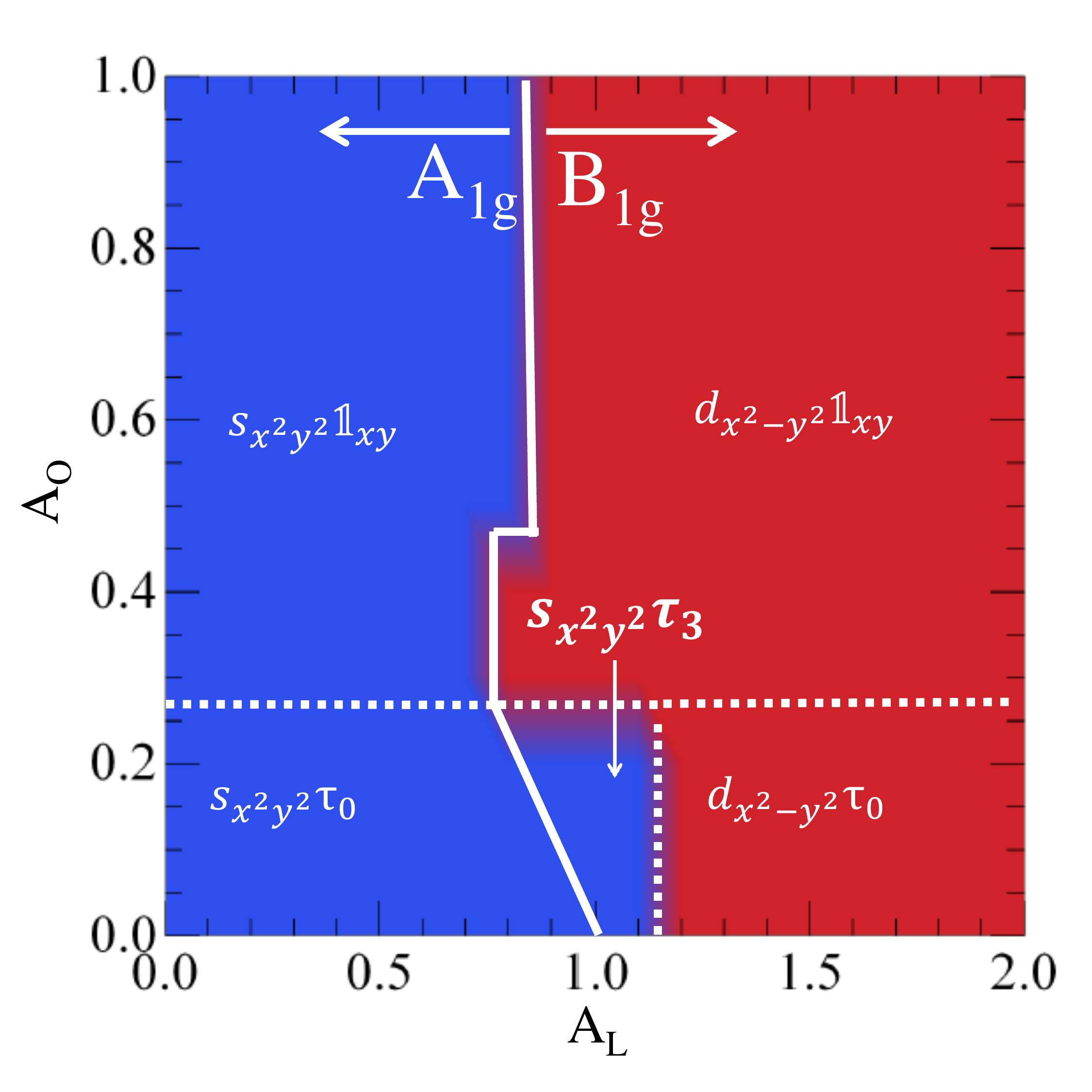}} \\[-4ex]
\subfloat[] {\includegraphics[width=0.8\columnwidth]{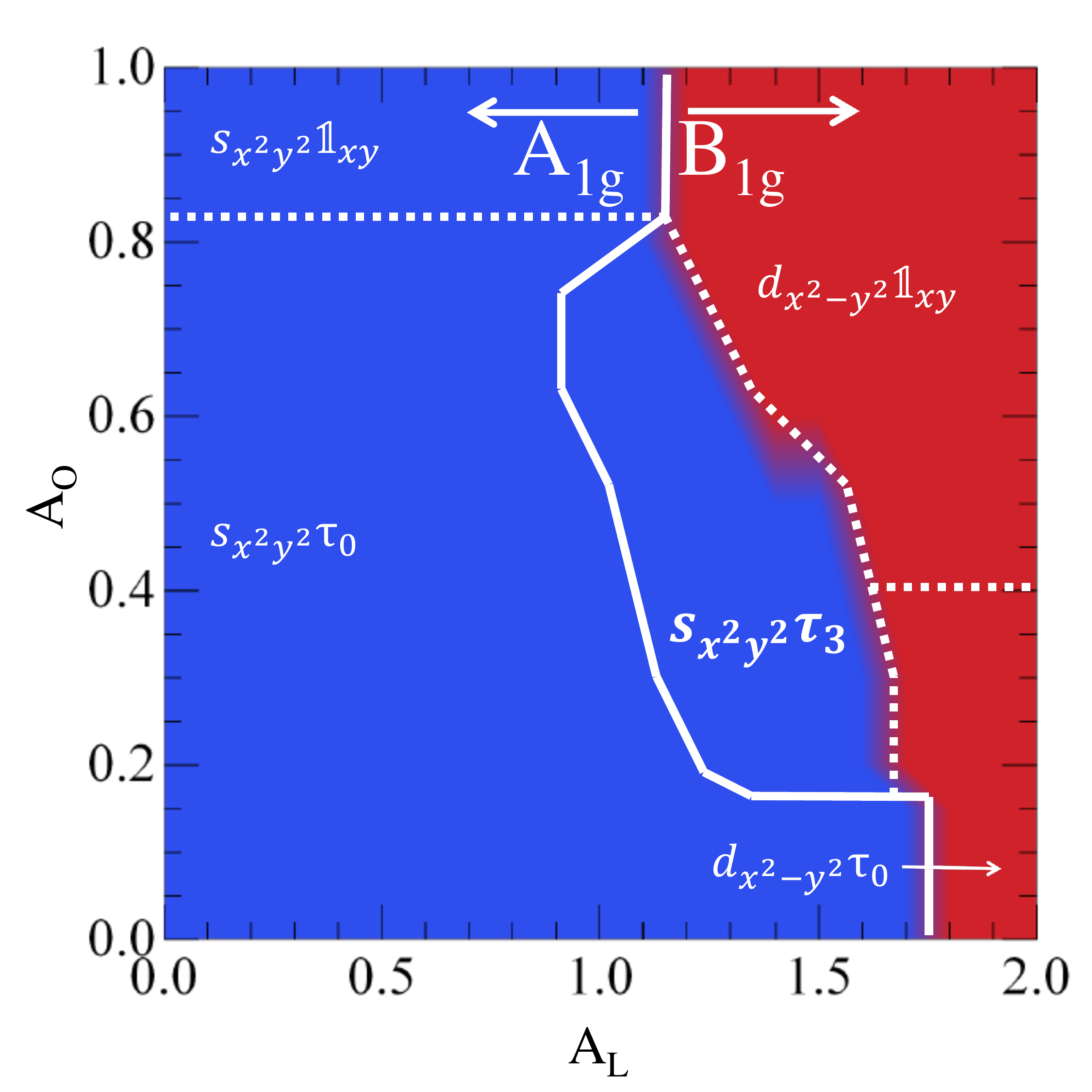}}
\caption{ Phase diagrams based on the leading pairing amplitudes given by self-consistent calculations with fixed $J_2=1$ and tight-binding parameters appropriate to (a) alkaline iron selenides, and (b) iron pnictides. For the tight-binding parameters used please consult Ref. \onlinecite{Yu_Nat_Comm:2013}. The blue shaded areas correspond to dominant pairing channels with an $s_{x^2y^2}$ form factor while the red shading covers those with a $d_{x^2-y^2}$ form factor. The continuous line separates regions where the pairing belongs to the $A_{1g}$ and the $B_{1g}$ representations respectively. The $1 \times 1$ matrix in the $d_{xy}$ subspace is represented by $\pmb{1}_{xy}$.  Note the presence of the intermediate $s_{x^2y^2}, B_{1g}$ pairing for $A_O < 1$, $A_L \geq 1$ in all cases.}
\label{Fig:Phase_diagrams_K_Fe_Se}
\end{figure}

\section{Orbital selectivity and the zero temperature superconducting state with dominant $s \times \tau_{3} $ pairing}
\label{Sec:Numerical_results}

We are now in position to discuss how the intermediate pairing state emerges in a range of parameters where the $s-$ and $d-$wave pairing channels are quasi-degenerate.
Within the 5-orbital $t-J_{1}-J_{2}$ model, we focus on the case with a kinetic part appropriate for
$ \textnormal{K}_{\textnormal{y}}\textnormal{Fe}_{\textnormal{2-x}}\textnormal{Se}_{\textnormal{2}} $ although similar behavior emerges in the iron pnictides and single-layer FeSe. All quantities are given in units of a half-bandwidth $D/2$,  which already incorporates a doping-dependent renormalization of the kinetic energy.

The $s_{x^2y^2} \times \tau_{3} $ pairing becomes dominant over a finite range of the tuning parameters ($A_L$ and $A_O$). For small $A_O$ and $A_L$ the leading pairing 
occurs in the $s_{x^2y^2} \times \tau_{0} , {A_{1g}}$  ("$s\pm$") channel. In this regime the strongest contribution comes from the NNN exchange coupling in the $d_{xz}$, $d_{yz}$ sector. 
By increasing  the $J_{1}-J_{2}$ ratio $A_L > 1$ i.e. moving along the horizontal axis, 
the $d_{x^2-y^2} \times \tau_{0}, B_{1g}$ in the $d_{xz}$, $d_{yz}$ subspace favored 
by a large NN coupling eventually takes over. These two limiting phases are consistent with the results 
obtained in Ref. \onlinecite{Yu_Zhu_Si:2014} where $A_O$ was set to unity for all values of $A_L$. 
For the intermediate values of the the $J_{1}-J_{2}$ anisotropy factor 
$0.9 \leq A_L \leq 1.1 $, near the regime where the above two pairing channels are quasi-degenerate,
the intermediate  $s_{x^2y^2} \times \tau_{3}, B_{1g}$ pairing state emerges as 
the dominant channel.  The phase diagram for the alkaline iron selenides is shown in Fig. \ref{Fig:Phase_diagrams_K_Fe_Se} (a). Note that the $s_{x^2y^2}\times \tau_{3} $ phase also persists for a finite range of $A_O<0.3$. Similar phase diagrams are obtained for the iron pnictides and single-layer FeSe  shown in Figs. \ref{Fig:Phase_diagrams_K_Fe_Se} (b) and \ref{Fig:Single_layer_Fe_Se} (Appendix) respectively. A typical dominant $s_{x^2y^2} \times \tau_3, B_{1g}$ pairing case is shown in Fig. \ref{Fig:Pairing_amplitude_A_O=0.5_A_L=1.3}  in the Appendix for a number of subleading symmetry-allowed channels \cite{Goswami_Nikolic_Si:EPL_2010} for alkaline iron selenide dispersion with fixed $J_2=1.5$, $A_O=0.3$ and varying $A_L$ (horizontal axis).

\noindent \begin{figure}[t!]
\subfloat[]{\includegraphics[width=0.9\columnwidth]{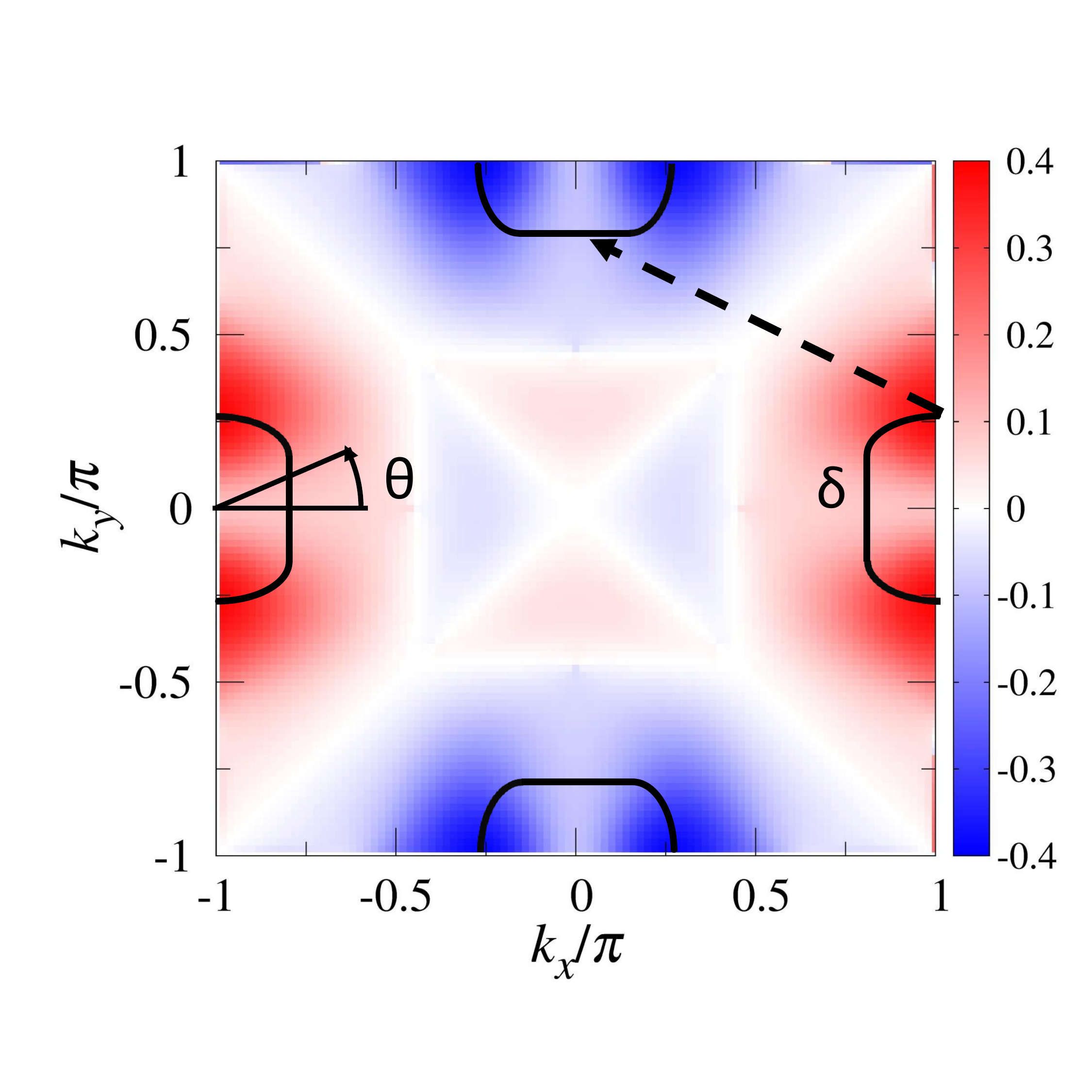}} \\[-4ex]
\subfloat[] {\includegraphics[width=0.9\columnwidth]{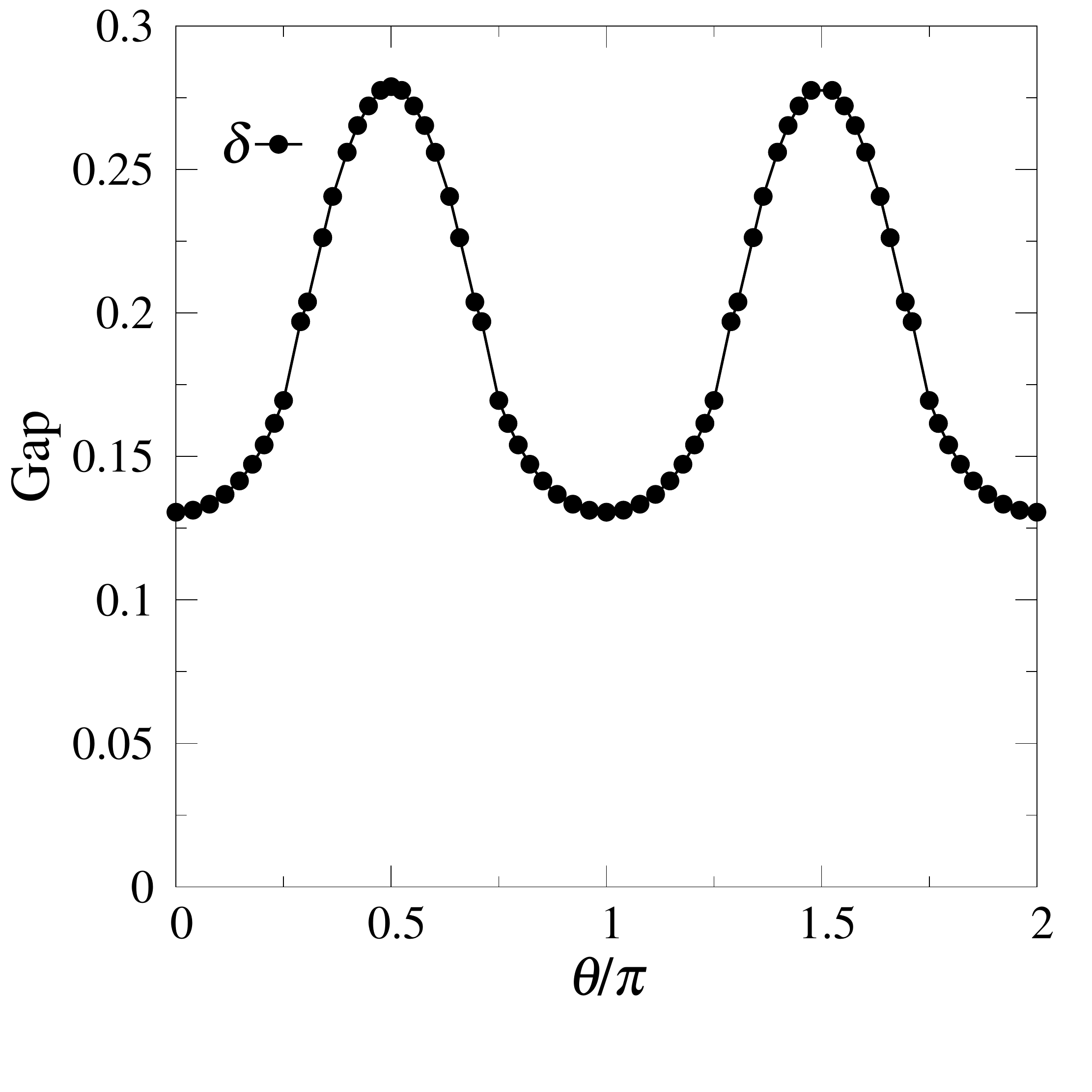}}
\caption{(a) The Fermi surface (solid line) and the real intra-band pairing for the band generating the $\delta$ pockets at the edge of the 1-Fe unit BZ for a dispersion typical of the alkaline iron selenides. Note the clear change in sign between pockets separated by the BZ diagonal. The dashed arrow indicates the $\pmb{q}=(\pi,\pi/2)$ wave-vector associated with the resonance in the spin spectrum found in experiment \cite{Dai:arxiv_2015}. (b) The size of the gap along the $\delta$ pocket. Both figures are for $J_2=1.5$, $A_O=0.3, A_L=0.9$ with dominant $s_{x^2y^2} \times \tau_{3}$ pairing. }
\label{Fig:K_FS_pair_sign}
\end{figure}

The Fermi surface and the sign of the band-diagonal pairing for the band generating the pockets around $(\pm \pi,0)$ ($\delta$) and $(0, \pm \pi)$ for the alkaline iron selenides with $J_2=1.5, A_O=0.3$ and $A_L=0.9$ is shown in Fig. \ref{Fig:K_FS_pair_sign} (a). The sign of the real pairing was obtained by projecting the full pairing matrix onto the band basis. The dominant contribution is from the $s_{x^2y^2} \times \tau_{3} $ component but all sub-leading channels (see Fig. \ref{Fig:Pairing_amplitude_A_O=0.5_A_L=1.3}) were included. The arbitrary phase which results from the global $U(1)$ symmetry breaking is subtracted from all components ensuring that the resulting pairing is real. We see that the band-diagonal pairing does indeed change sign between the two pockets at the edge of the BZ zone. In Fig. \ref{Fig:K_FS_pair_sign} (b) we show the gap at the Fermi surface as a function of winding angle $\theta$. The figure clearly illustrates the node-less dispersion as t
 he gap is finite for all $\theta$.

\noindent \begin{figure}[]
\subfloat[]{\includegraphics[width=0.9\columnwidth]{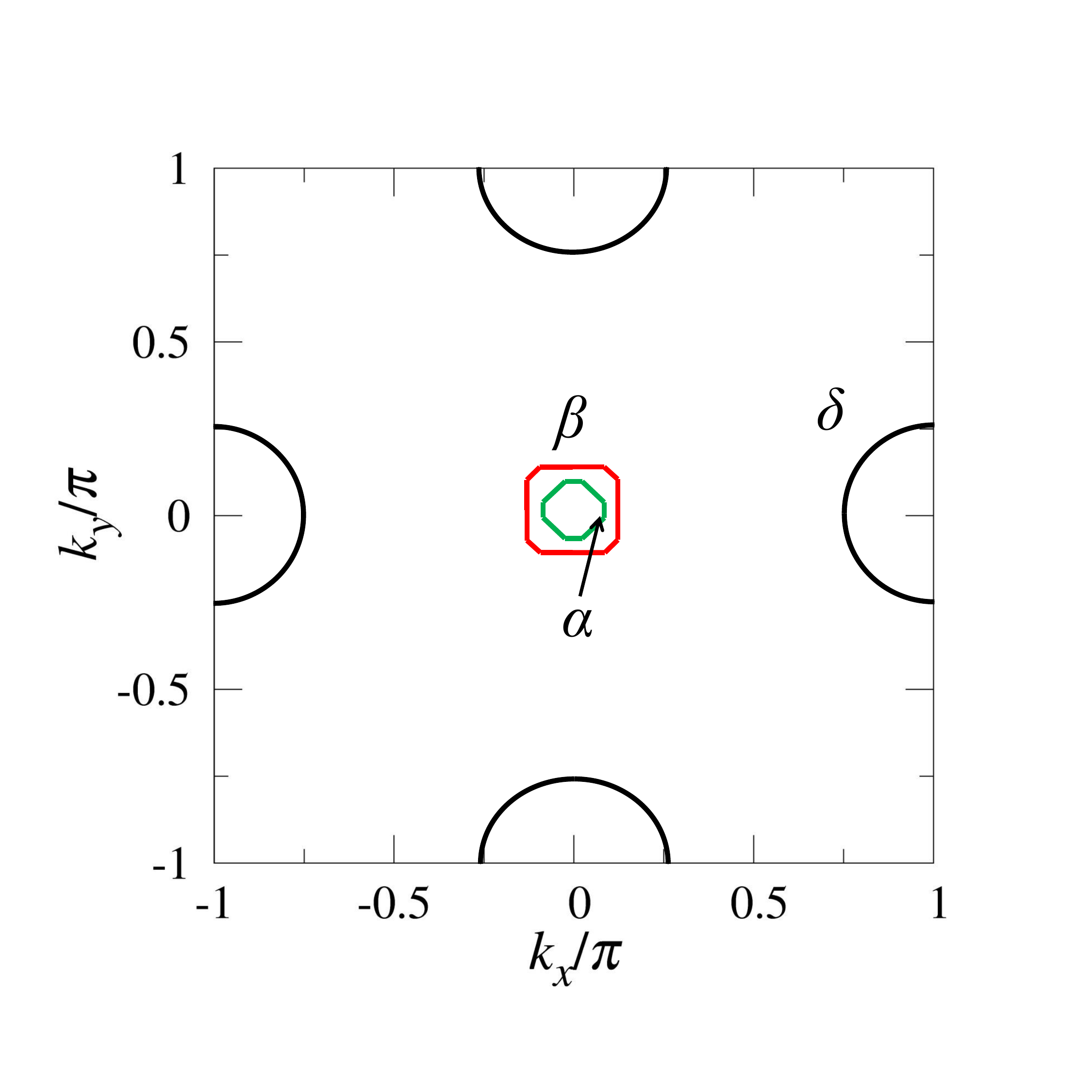}} \\[-4ex]
\subfloat[]{\includegraphics[width=0.9\columnwidth]{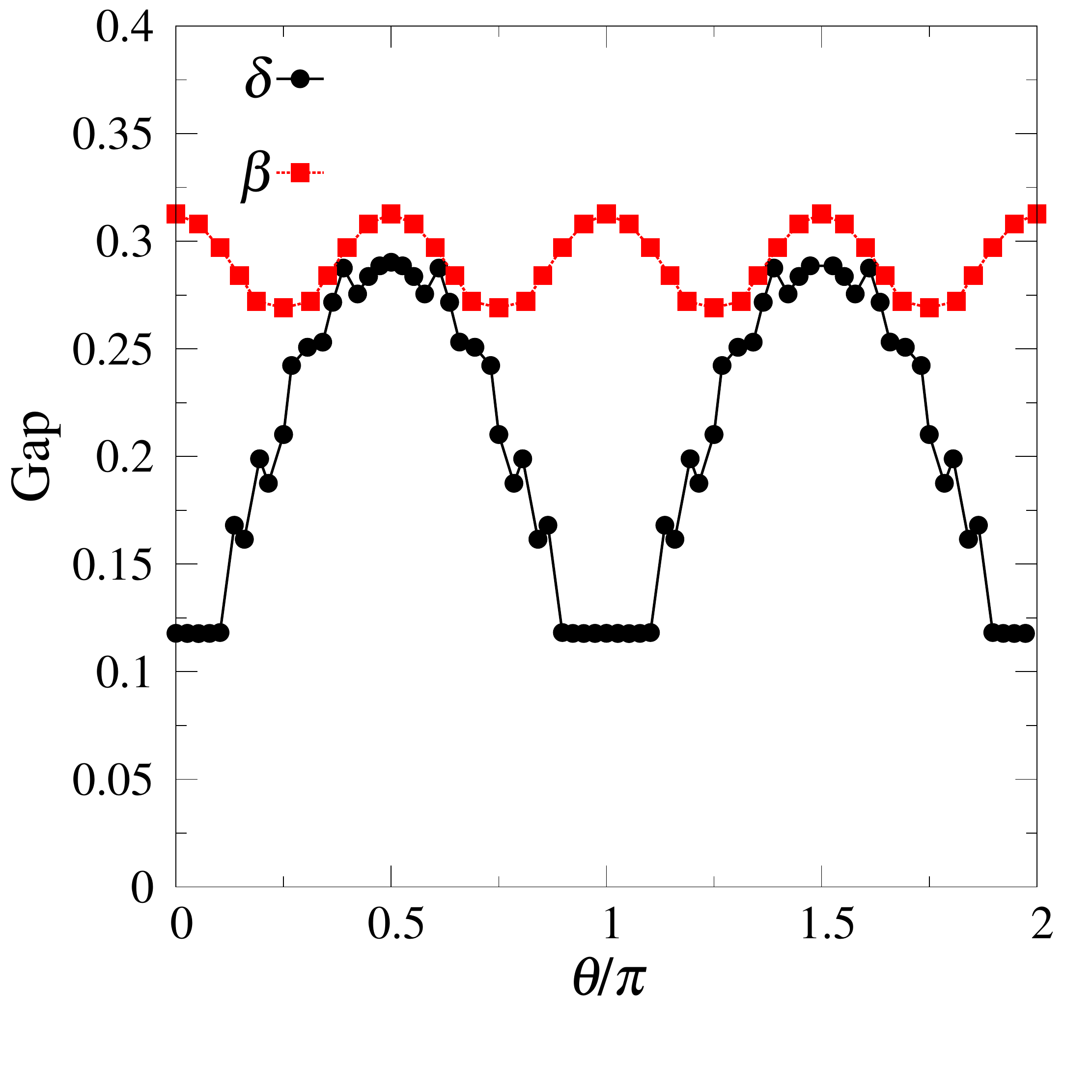}}
\caption{ (a) Fermi surface for the iron pnictides which includes hole pockets with dominant $s_{x^2y^2} \times \tau_{3}, B_{1g}$ close to $\Gamma$ for $J_2=1, A_L=1.3, A_O=0.5 $. For the tight-binding parameters used please consult Ref. \onlinecite{Yu_Nat_Comm:2013}. (b) The gaps along the $\beta$ and $\delta$ pockets close to the center and edge of the 1-Fe BZ. A similar gap forms around the $\alpha$ pocket.}
\label{Fig:A_FS_gap}
\end{figure}

The free-particle dispersion considered here does not produce any Fermi pockets close to $\Gamma$ in the 1-Fe BZ. This is in contrast to ARPES experiments on $ \textnormal{K}_{\textnormal{y}}\textnormal{Fe}_{\textnormal{2-x}}\textnormal{Se}_{\textnormal{2}} $ \cite{Zhang_et_al:2011} which show a small electron pocket near $\Gamma$. Because this small electron pocket has very small spectral weight, it is to be expected that even if such a pocket were included, the dominant $s_{x^2y^2} \times \tau_{3} $ pairing will still arise; moreover, the gap on this Fermi pocket will be node-less as discussed in the two-orbital case. To substantiate this, we consider the results for the iron pnictides class, which do have significant ({\it albeit} hole) Fermi pockets
 at the zone center yet exhibit a full gap. In Figs. \ref{Fig:A_FS_gap} (a), (b) we show the Fermi surface and the gaps as functions of winding angle $\theta$ for $A_O=0.5$ and $A_L=1.3$ corresponding to a dominant $s_{x^2y^2} \times \tau_{3} $ pairing. The gap along $\beta$ is finite and exhibits an anisotropy consistent with the two orbital results in Eq.\ref{Eq:Balian_Wert_dispersion}. In the latter case, at winding angle $\theta=0$, $\text{sin}\phi=0$ and the spectrum has a minimum/maximum gap for $E_{+/-}$. As $\theta$ is increased the $\left| \overrightarrow{B}(\pmb{k}) \times \overrightarrow{d}(\pmb{k}) \right|^2$ term increases reaching a maximum at $\theta=\pi/4$. Here the gap is maximum/minimum for $E_{+/-}$. This is consistent with the anisotropy in the gap shown in Fig. \ref{Fig:A_FS_gap}. We stress that the fully gapped dispersion is \emph{not} the result of a sub-leading s-wave $A_{1g}$ channel. As in the alkaline iron selenide case, all $A_{1g}$ pairing functio
 ns are strongly suppressed $(\text{O}(10^{-3})$) at this point (See Fig. \ref{Fig:Pairing_amplitude_A_O=0.5_A_L=1.3} in the Appendix).

 \noindent \begin{figure}[!htb]
\includegraphics[width=0.9\columnwidth]{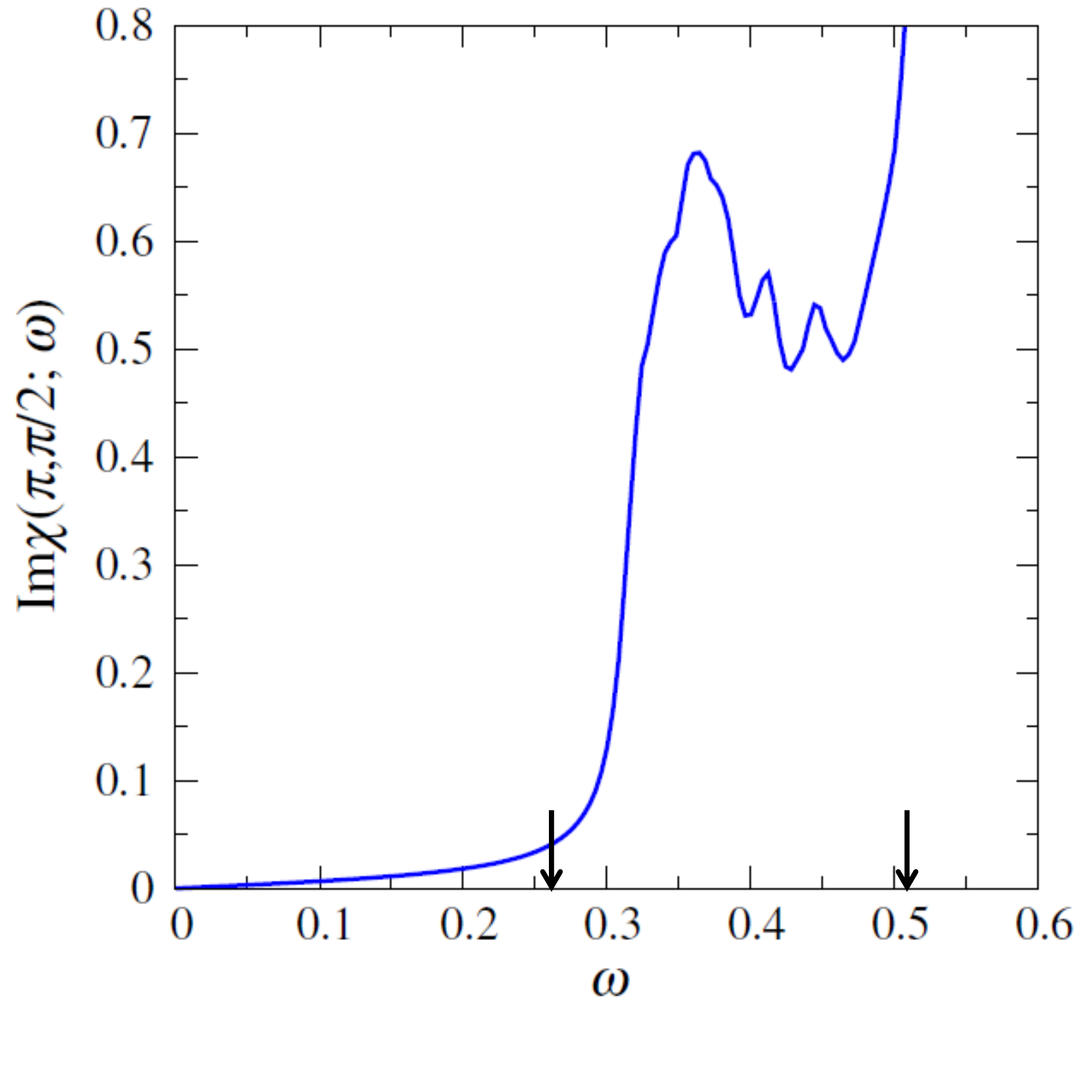}
\caption{The imaginary part of the RPA susceptibility (Eq. \ref{Eq:RPA_suscept}) for the alkaline iron selenides at wave-vector $\pmb{q}=(\pi, \pi/2)$, for a dominant $s_{x^2y^2} \times \tau_{3}$ pairing. The arrows show twice the minimum and maximum gaps (see Fig. \ref{Fig:K_FS_pair_sign} (b)). There is a sharp feature ar $\omega \approx 0.36$ within the bounds of twice the effective gap and below the p-h threshold of roughly 0.41 associated with this wavevector.}
\label{Fig:K_spin_spectrum}
\end{figure}

The full gap and the sign change provide evidence that with strong orbital selectivity the $s_{x^2y^2} \times \tau_{3}$ pairing in a realistic five-orbital $t-J_{1}-J_{2}$ model has a behavior very similar to the two-orbital case and that it can be considered a viable candidate for the alkaline iron selenide superconductors.

In Fig. \ref{Fig:K_spin_spectrum} we show the spin-excitation spectrum calculated for the alkaline iron selenides with dominant $s_{x^2y^2} \times \tau_{3}, B_{1g}$  pairing at wave-vector $\pmb{q}=(\pi, \pi/2)$ for $J_2=1.5$. We note the complicated frequency behavior which can be traced to the anisotropy in the effective gap affecting both the coherence factors and the position of minimum in quasi-particle energy. We show the minimum and maximum p-h thresholds corresponding to twice the minimum and twice the maximum gaps. As suggested by Figs. \ref{Fig:K_FS_pair_sign} (a) and (b), states connected by $\pmb{q}=(\pi, \pi/2)$ would correspond to a p-h threshold given roughly by the sum of the minimum and maximum gap $\approx 0.41$. A sharp feature appears below this threshold at $\omega \approx 0.36$, confirming the existence of the resonance for $\pmb{q}=(\pi, \pi/2)$ as found in experiments on the alkaline iron selenides \cite{Dai:arxiv_2015}.

\section{Conclusion}
\label{Sec:Conclusion}

We have shown that, through orbital selectivity, an intermediate pairing state emerges in the regime where the conventional $s-$ and $d-$wave pairing channels are quasi-degenerate. This superconducting state is energetically competitive, as illustrated by our calculations in a five-orbital $t-J_{1}-J_{2}$ model with orbital-selective exchange couplings.

This intermediate pairing state inherits aspects of the properties of {\it both} the conventional $s-$ and $d-$wave pairing channels. As we have explicitly illustrated in the case of  $d_{xz}, d_{yz}$ orbital subspace, 
this orbital-selective $s \times \tau_{3}, B_{1g}$  pairing state has the
$s$-wave form factor but also has a $B_{1g}$ symmetry. Going beyond this two-orbital subspace, the contributions of the other orbitals mix into the pairing function. Nonetheless, the pairing state still incorporates some of the properties of both the $s-$ and $d-$wave pairing states.

For the iron-based superconductors, this intermediate state is of considerable phenomenological interest. In particular, it has the salient properties observed in the alkaline iron selenides. These properties include seemingly contradictory aspects. The single-particle excitations are fully gapped, as observed in ARPES experiments. At the same time, the pairing function changes sign across the electron Fermi surfaces at the BZ boundary, as indicated by the resonance peak near $(\pi,\pi/2)$ in the inelastic neutron scattering experiments.

More generally, a conventional means of relieving quasi-degenerate $s-$ and $d-$wave pairing states is to linearly superpose the two into an $s+id$ state. This state, breaking the time-reversal symmetry, would be stabilized at temperatures sufficiently below the superconducting transition temperature. The mechanism advanced here preserves the time-reversal symmetry, and represents a new means to relieve the quasi-degeneracy through the development of orbital selectivity. 

{\it Acknowledgements.~} 
This work has been supported by the NSF Grant No. DMR-1309531, the Robert A.\ Welch Foundation
Grant No.\ C-1411 (E.M.N. \& Q.S.).
R.Y. was partially supported by the National Science Foundation of China
Grant number 11374361, and the Fundamental Research Funds for the
Central Universities and the Research Funds of Renmin University
of China.
All of us acknowledge the support provided in part by the
NSF Grant  No. NSF PHY11-25915 at KITP, UCSB, for our participation in the Fall 2014
program on ``Magnetism, Bad Metals and Superconductivity: Iron Pnictides and Beyond".
Q.S.\ also acknowledges the hospitality
of the Institute of Physics of Chinese Academy of Sciences.

\appendix
\section{Supplementary Material}
\label{Sec:Appendix}

The components of the dispersion part of the two-orbital Hamiltonian discussed in Sec. \ref{Sec:Two_orbital} are given by 

\noindent \begin{align}
\xi_{\pmb{k}+} = & -(t_1+t_2)(\cos k_x+\cos k_y) \notag \\
& -4t_3\cos k_x \cos k_y, \\
\xi_{\pmb{k}-}= & -(t_1-t_2)(\cos k_x-\cos k_y), \\
\xi_{\pmb{k}xy}= & -4t_4\sin k_x \sin k_y,
\end{align}

\noindent where $t_1$,$t_2$ and $t_3$ are tight-binding parameters. 
The parameters of the two-orbital and five-orbital models are those given in
Ref. \onlinecite{Yu_Nat_Comm:2013}. The free-band dispersion is given by

\noindent \begin{equation}
\label{Eq:Free-band}
\epsilon_{\pm}(\pmb{k})= \xi_{+}(\pmb{k}) \pm \sqrt{\xi^2_{-}(\pmb{k})+ \xi^2_{xy}(\pmb{k})}.
\end{equation}

In the single-band BCS case, the leading contribution to the dynamical spin susceptibility (see Eq. \ref{Eq:RPA_suscept} for the multi-orbital case) depends \cite{Eschrig:Adv_Phys_2006, Fong_et_al:PRL_1995}  on terms like

\noindent \begin{widetext}
\begin{align}
\label{Eq:BCS_RPA_suscept}
\chi_{0}(\pmb{q},\omega)=\frac{1}{N}\sum_{\pmb{k}} \bigg[ \frac{1}{2} \bigg( 1- \frac{\epsilon_{\pmb{k}+\pmb{q}} \epsilon_{\pmb{k}} + \Delta_{\pmb{k}+\pmb{q}} \Delta_{\pmb{k}} }{E_{ \pmb{k} + \pmb{q} } E_{\pmb{q}} } \bigg) \frac{ f( E_{ \pmb{k} + \pmb{q} }) + f(E_{\pmb{k}}) -1 }{ \omega - ( E_{ \pmb{k} + \pmb{q} }+ E_{ \pmb{k} } ) + i0^{+} } + ... \bigg] ,
\end{align}
\end{widetext}

\noindent where $\epsilon$'s and $E$'s are the free particle and the BdG quasi-particle dispersions respectively. The existence of a sharp feature in the RPA spectrum below the particle-hole threshold (given roughly by twice the characteristic gap magnitude $2\Delta$) is related to the sign of the $\Delta_{\pmb{k}+\pmb{q}} \Delta_{\pmb{k}}$ term in the coherence factor in Eq. \ref{Eq:BCS_RPA_suscept}.  Close to the Fermi surface, when the sign is positive, the coherence factor suppresses the real part of $\chi_{0}(\pmb{q},\omega)$ and consequently, inhibits the appearance of a resonance. By contrast, when $\Delta_{\pmb{k}+\pmb{q}}$ and $\Delta_{\pmb{k}}$ have opposite signs, the resonance can form. For multi-band systems with non-trivial gap dependence, the situation is obviously more complicated.  Nonetheless, the occurrence of the spin resonance is typically still connected with a sign change in the gap function.

\noindent \begin{figure}[!htb]
\includegraphics[width=0.9\columnwidth]{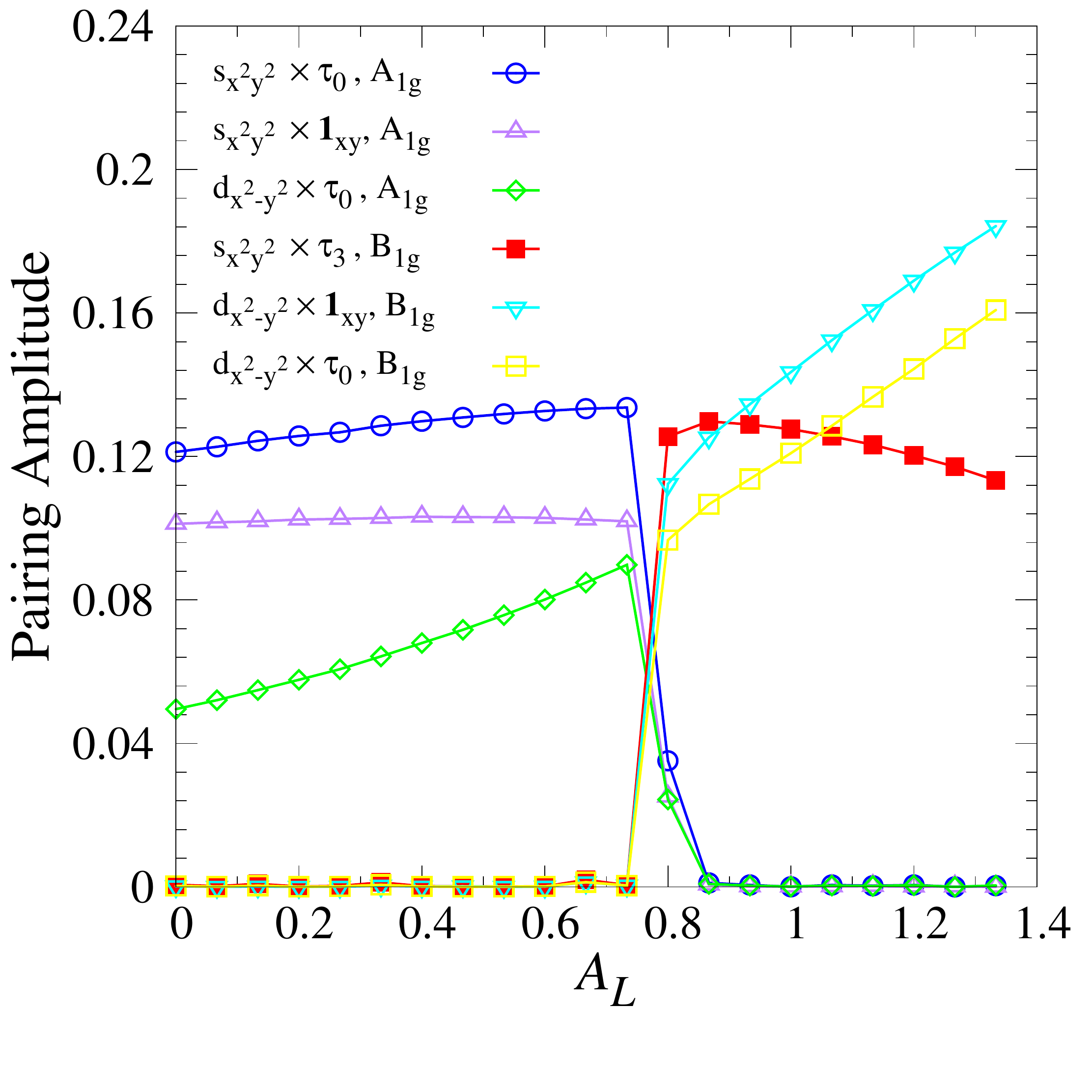}
\caption{Leading pairing amplitudes (vertical axis) for a dispersion typical of alkaline iron selenides for fixed  $J_2=1.5, A_O=0.3$ and varying NN-NNN ratio $A_L$ (horizontal axis). The $\tau$ label indicates a dominant $d_{xz}, d_{yz}$ sector while $ \mathbf{1}_{xy}$ marks a $d_{xy}$ dominant pairing. For $ 0.8 \leq A_L \leq 0.94 $ the leading pairing is in the $s_{x^2y^2}\times \tau_{3}, {B1g}$ channel shown in dark filled squares. Note that the reduced parameter space for the $s_{x^2y^2}\times \tau_{3}, {B1g}$ is due to the proximity to the phase boundary and for lower values of $A_O$ the range over which this pairing leads is increased.}
\label{Fig:Pairing_amplitude_A_O=0.5_A_L=1.3}
\end{figure}

\noindent \begin{figure}
\includegraphics[width=0.8\columnwidth]{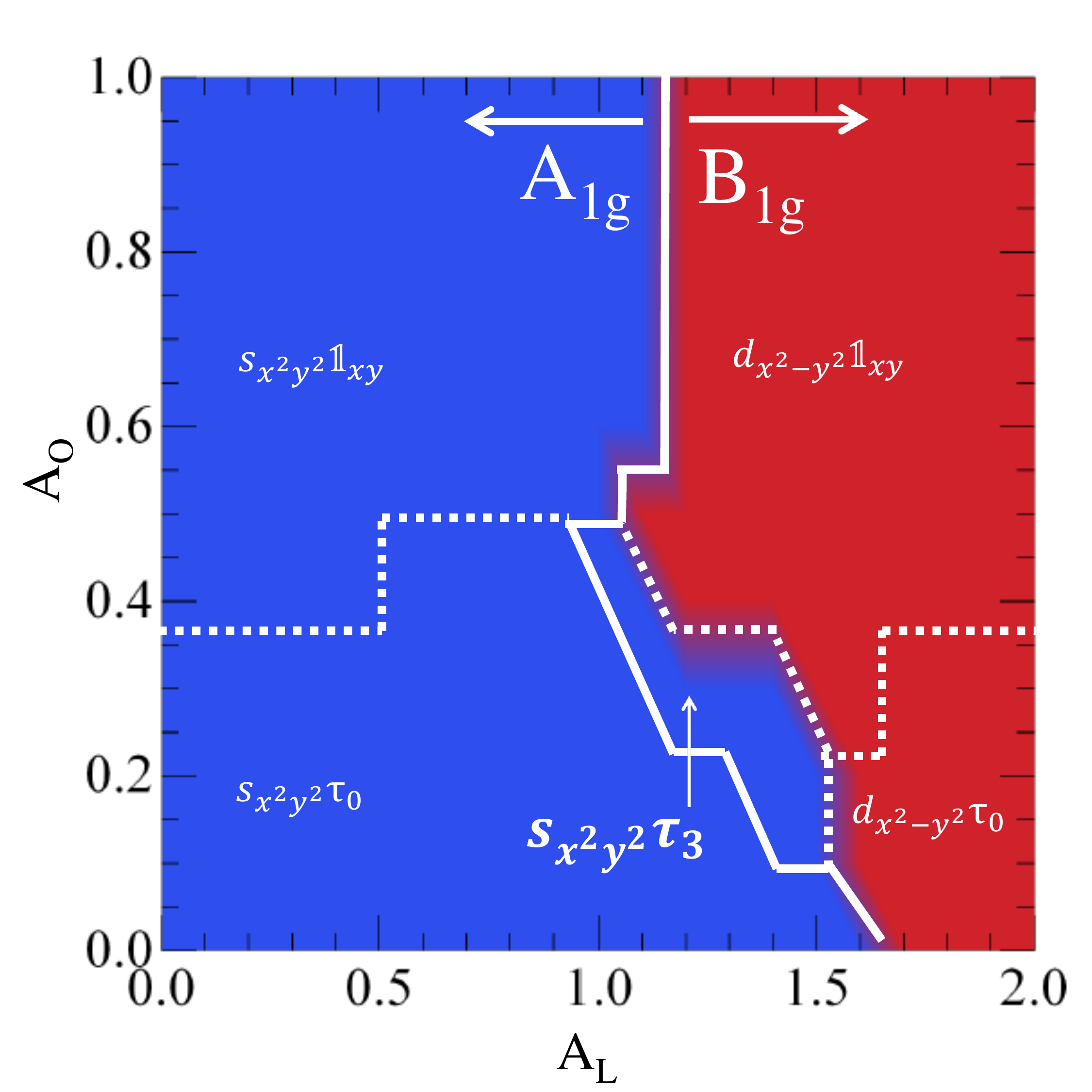}
\caption{ Phase diagram based on the leading pairing amplitudes given by self-consistent calculations using tight-binding parameters appropriate to single-layer FeSe. For the tight-binding parameters used please consult Ref. \onlinecite{Yu_Nat_Comm:2013}. The blue shaded areas correspond to dominant pairing channels with an $s_{x^2y^2}$ form factor while the red shading covers those with a $d_{x^2-y^2}$ form factor. The continuous line separates regions where the pairing belongs to the $A_{1g}$ and the $B_{1g}$ representations respectively. The $1 \times 1$ matrix in the $d_{xy}$ subspace is represented by $\pmb{1}_{xy}$.}
\label{Fig:Single_layer_Fe_Se}
\end{figure}

\end{document}